\documentclass[12pt]{article}
\pdfoutput=1
\usepackage{amssymb}
\usepackage{amsmath}
\usepackage{amsfonts}
\usepackage{graphicx}
\usepackage{mathrsfs}
\usepackage{comment}
\usepackage{cite}

\begin{document}

\renewcommand{\figurename}{Fig.}
\renewcommand{\tablename}{Table.}
\newcommand{\Slash}[1]{{\ooalign{\hfil#1\hfil\crcr\raise.167ex\hbox{/}}}}
\newcommand{\bra}[1]{ \langle {#1} | }
\newcommand{\ket}[1]{ | {#1} \rangle }
\newcommand{\beq}{\begin{equation}}  \newcommand{\eeq}{\end{equation}}
\newcommand{\bef}{\begin{figure}}  \newcommand{\eef}{\end{figure}}
\newcommand{\bec}{\begin{center}}  \newcommand{\eec}{\end{center}}
\newcommand{\non}{\nonumber}  \newcommand{\eqn}[1]{\begin{equation} {#1}\end{equation}}
\newcommand{\laq}[1]{\label{eq:#1}}  
\newcommand{\dd}[1]{{d \o d{#1}}}
\newcommand{\Eq}[1]{Eq.(\ref{eq:#1})}
\newcommand{\Eqs}[1]{Eqs.(\ref{eq:#1})}
\newcommand{\eq}[1]{(\ref{eq:#1})}
\newcommand{\Sec}[1]{Sec.\ref{chap:#1}}
\newcommand{\ab}[1]{\left|{#1}\right|}
\newcommand{\vev}[1]{ \left\langle {#1} \right\rangle }
\newcommand{\bs}[1]{ {\boldsymbol {#1}} }
\newcommand{\lac}[1]{\label{chap:#1}}
\newcommand{\SU}[1]{{\rm SU{#1} } }
\newcommand{\SO}[1]{{\rm SO{#1}} }
\def\({\left(}
\def\){\right)}
\def\dt{{d \o dt}}
\def\diag{\mathop{\rm diag}\nolimits}
\def\Spin{\mathop{\rm Spin}}
\def\O{\mathcal{O}}
\def\U{\mathop{\rm U}}
\def\Sp{\mathop{\rm Sp}}
\def\SL{\mathop{\rm SL}}
\def\tr{\mathop{\rm tr}}
\def\ebq{\end{equation} \begin{equation}}
\newcommand{\OR}{~{\rm or}~}
\newcommand{\AND}{~{\rm and}~}
\newcommand{\EV}{ {\rm \, eV} }
\newcommand{\KEV}{ {\rm \, keV} }
\newcommand{\MEV}{ {\rm \, MeV} }
\newcommand{\GEV}{ {\rm \, GeV} }
\newcommand{\TEV}{ {\rm \, TeV} }
\def\o{\over}
\def\a{\alpha}
\def\b{\beta}
\def\c{\varepsilon}
\def\d{\delta}
\def\e{\epsilon}
\def\f{\phi}
\def\g{\gamma}
\def\h{\theta}
\def\k{\kappa}
\def\l{\lambda}
\def\m{\mu}
\def\n{\nu}
\def\p{\psi}
\def\q{\partial}
\def\r{\rho}
\def\s{\sigma}
\def\t{\tau}
\def\u{\upsilon}
\def\v{\varphi}
\def\w{\omega}
\def\x{\xi}
\def\y{\eta}
\def\z{\zeta}
\def\D{\Delta}
\def\G{\Gamma}
\def\H{\Theta}
\def\L{\Lambda}
\def\F{\Phi}
\def\P{\Psi}
\def\S{\Sigma}
\def\me{\mathrm e}
\def\ol{\overline}
\def\tl{\tilde}
\def\*{\dagger}

\newcommand{\rem}[1]{{$\spadesuit$\bf #1$\spadesuit$}}

\begin{center}

\hfill 

\vspace{1cm}

{\Large\bf  Radiative lepton mass and muon $g-2$ with
suppressed lepton flavor and CP violations}
\vspace{1.5cm}

{\bf Wen Yin }
\vspace{1.5cm}

{Department of Physics, Tohoku University,  
Sendai, Miyagi 980-8578, Japan \\} 
{Department of Physics, Faculty of Science, The University of Tokyo,  Bunkyo-ku, Tokyo 113-0033, Japan}
\vspace{1.5cm}

\vspace{12pt}
\vspace{1.5cm}

\date{\today $\vphantom{\bigg|_{\bigg|}^|}$}

\abstract{
{ The recent experimental status, including the confirmation of the muon $g-2$ anomaly at Fermilab, indicates a Beyond Standard Model (BSM) satisfying the following properties: 
1) it enhances the $g-2$ 2) suppresses flavor violations, such as $\mu \to e \g$, 3) suppresses CP violations, such as the electron electric dipole moment (EDM). 
In this letter, I show that if the masses of heavy leptons are generated radiatively, 
the eigenbasis of the mass matrix and higher dimensional photon operators can be automatically aligned. 
As a result, the muon $g-2$ is enhanced but the EDM of the electron and $\mu \to e \gamma$ rate are suppressed. 
Phenomenology and applications of the mechanism to the B-physics anomalies are argued. 
}

 }

\end{center}
\clearpage

\setcounter{page}{1}
\setcounter{footnote}{0}

\setcounter{footnote}{0}
\section{Introduction}

The anomalous muon magnetic moment ($g-2$) observed by the  Brookhaven result \cite{Bennett:2006fi} has been confirmed by Fermilab~\cite{Fermilab,PhysRevLett.126.141801}. The combined $g-2$ anomaly is 
\begin{align}
\laq{g2m}
\D a_\mu &= a_\mu^{\rm EXP} - a_\mu^{\rm SM} = (25.1 \pm 5.9) \times 10^{-10},
\end{align}
where $a_\mu^{\rm SM}$ is the standard model (SM) prediction of the $(g-2)/2$ (See also Refs.~\cite{Roberts:2010cj, Davier:2017zfy,Keshavarzi:2018mgv, Keshavarzi:2019abf, Borsanyi:2020mff, Aoyama:2020ynm, Chao:2021tvp})\footnote{See also Ref.\,\cite{Keshavarzi:2019abf}  for the lattice result of the muon $g-2$, which is smaller than that from the $R$-ratio approach. The explanation of the $g-2$ within the SM is an important topic but we may need further checks~\cite{Crivellin:2020zul, Keshavarzi:2020bfy}.}.
$a_\mu^{\rm EXP}$ is the combined updated experimental value (See also Refs.~\cite{Bennett:2006fi, Roberts:2010cj,Parker:2018vye, Hanneke:2008tm,Hanneke:2010au}.) 
The deviation is around $4.2 \sigma$. (If we adopt the R-ratio analysis in \cite{Keshavarzi:2019abf} it is at $4.5\s$ level.) 
There are also anomalies relevant to B-physics, especially the lepton non-universality \cite{Aaij:2021vac, talk}. For concreteness, we mainly focus on the $g-2$, while our mechanism can also apply to models for explaining the lepton non-universality as will be discussed in the last section.

The $g-2$ result strongly implies beyond SM (BSM) particle(s) coupled to the SM leptons, especially the muon. 
The results, together with the non-observation of the charged lepton EDM and the charged lepton flavor violation, 
 provide a hint to the flavor and CP structure of the BSM. 
On the other hand, in collider experiments, there is no evidence of the direct productions of the BSM particles, which may suggest the BSM particles are heavy. 
If the BSM particles are heavier than TeV, the sizable muon $g-2$ should be generated via a process without a chirality suppression.
This implies that the SM lepton Yukawa couplings are likely to obtain a huge quantum effect from a similar diagram for the $g-2$. 
Therefore, in this Letter, I would like to study the radiative generation of the lepton Yukawa coupling in the context of the muon $g-2$. See e.g. Refs.\,\cite{Borzumati:1999sp,Crivellin:2010ty, Endo:2019bcj} for the SUSY theory relating the lepton $g-2$ and the fermion mass hierarchy.

In particular, I show that the muon $g-2$ can be easily explained when the muon mass is radiatively induced. 
In addition, 
the $\mu \to e\g $ rate and the electron EDM are suppressed since the induced higher dimensional photon coupling can be automatically aligned to the muon eigenstate of the lepton mass matrix.

\section{  Heavy lepton from radiative correction.}
\lac{1}
Let me give a general discussion to present the idea. 
For simplicity let us consider leptons with two flavors. Later they will be identified as electron and muon. 
I will also assume that the BSM particles couple to only one combination of the leptons without flavor/CP symmetry. This is the case if the ``family" of the BSM particles is $1$.

The relevant interacting Lagrangian at the cutoff scale, $\mu\sim \L_{\rm cutoff},$ with normalized kinetic terms is given as 
\beq
{\cal L}_{\rm int}\supset -\sum_{ij}y_{ij} H_0 \bar{L}_i e_j-{\cal L}_{\rm BSM}[L_2, e_2, \O]
\eeq 
where $L_i$ ($e_j$) are the left (right) handed charged leptons, $H_0=v+\frac{h}{\sqrt{2}}$ the Higgs VEV plus the Higgs boson, $y_{ij}$ Yukawa matrix, 
 and $\O$ a set of heavy BSM particles of mass scale $\L$.\footnote{It is straightforward to add $\SU(2)$ partners for the gauge invariance. }
${\cal L}_{\rm BSM}[L_2,E_2, \O]$ is a BSM interaction term.  Here  we define the interacting leptons to have an index $2$ without loss of generality. 
In this basis the Yukawa matrix $y_{ij}$ naturally has a general flavor structure and 
I assume  
\beq
y_{ij}=\O(y)
\eeq
independent of $i, j$, where $y$ represents a typical size of the Yukawa coupling. This is taken to be 
\beq
y\ll1. 
\eeq 

By integrating out the heavy BSM particles, $\O$, 
the relevant terms in the effective Lagrangian at the scale $\mu\sim \L$ are
\begin{align}
{\cal L}_{\rm eff}\sim  &\bar{L}_1\partial_\mu \g^\m L_1+Z_L \bar{L}_2\partial_\mu \g^\m L_2 +\bar{e}_1\partial_\mu \g^\m e_1+Z_R \bar{e}_2\partial_\mu \g^\m e_2  \\
&- y_{ij} H_0 L_i e_j - \d Y  H_0 L_2 e_2\\
& - \frac{H_0}{\L^2} (\Re{[C_{\g}]} L_2 \sigma_{\mu\nu} F_{\mu \nu} e_2 +i \Im{[C_\g]}  L_2 \g_5 \sigma_{\mu\nu} {F}_{\mu \nu} e_2 )
\end{align}
where $Z_{L,R}$ are the wave function renormalization, $\d Y$ radiatively generated Yukawa terms,  and $C_\g/\L^2$ higher dimensional coupling in chiral representation.
Here $\s_{\mu \nu}\equiv i[\g_\m ,\g_\n]/2, $ and $F^{\m\n}$ is the field strength of photon. 
We have neglected the loop effects of order $y^2$. 
Also weak boson loop and Higgs wave function renormalization are neglected since the effects are either chirality suppressed or flavor blind. 

Significant corrections are all for the lepton with the index $``2"$. 
In particular the Yukawa matrix in the effective theory by assuming \beq 
\laq{radiative}
|\d Y|\gg |y_{22}|\eeq and neglecting $y_{22}$ is
\begin{align}
 Y_{ij}^{\rm eff} \propto \left(
    \begin{array}{cc}
      y_{11} & y_{12}/\sqrt{Z_R}   \\
      y_{21}/\sqrt{Z_L} & \d Y/\sqrt{Z_L Z_R}  \\
    \end{array}
  \right)
\end{align}
with normalized kinetic terms. 
Here we can take $y_{11}$, $\d Y$, to be real positive parameters by  performing a field redefinition via chiral rotations of the lepton fields. In this case $y_{12} y_{21}$ is in general a complex parameter. 
We can easily diagonalize this matrix to get the mass eigenstates of the muon and electron as
\beq
L_{\a} = U_L L, \AND e_\b= U_R R
\eeq 
where $U_{L,R}$ are unitary matrices satisfying
\begin{align}
(U_L)_{\alpha i}\approx \left(
    \begin{array}{cc}
1& -\sqrt{Z_R}\frac{y^*_{12}}{dY}  \\
     \sqrt{Z_R}\frac{y_{12}}{dY} & 1  \\
    \end{array}\)
    \\
    (U_R)_{\alpha i}\approx \left(
    \begin{array}{cc}
1&  -    \sqrt{Z_L}\frac{y^*_{21}}{dY}  \\
     \sqrt{Z_L}\frac{y_{21}}{dY} & 1  \\
    \end{array}\).
\end{align}
$\a=e, \mu$ represents the Yukawa/mass eigenstate. 
The Eigen masses are found to be
\beq
m_e= y_{11} v+\O(y^2/\d Y) , \AND m_\mu =\frac{ \d Y v}{\sqrt{Z_L}\sqrt{Z_R}}+ \O( y^2/\d Y).
\eeq

 We can now see that  \Eq{radiative} is the condition that the muon mass is radiatively generated. 
The effective Lagrangian of the SM has additional terms\footnote{In general we may also have terms of $\bar{L} e \bar{e}e$ contributing to $\mu \to eee$. In our scenario, this process is highly suppressed.  }
\beq
{\cal L}_{\rm eff}=- \frac{v}{\L^2} \(\Re{[\hat{C}_{\g}]}_{\a \b} L^\a \sigma_{\mu\nu} F_{\mu \nu} e^\b +i \Im{[\hat{C}_{\g}]}_{\a \b} L^\a \g_5 \sigma_{\mu\nu} F_{\mu \nu} e^\b \),
\eeq
in the EW breaking phase. 
The coefficients satisfy
\begin{align}
\sqrt{Z_L Z_R} \times \frac{\hat{C}_\g}{C_\g}\approx  \left(
    \begin{array}{cc}
\frac{y_{21} y_{12} \sqrt{Z_L Z_R}}{ \d Y^2}  & -\frac{y_{12}\sqrt{Z_R}}{\d Y}   \\
     -\frac{y_{21}\sqrt{Z_L}}{\d Y } & 1  \\
    \end{array}\)= \left(
    \begin{array}{cc}
\frac{y_{12}y_{21}}{y^2_{11}}\frac{ m_e^2 }{\sqrt{Z_R Z_L} m_\mu^2} &-\frac{y_{12}}{y_{11}} \frac{m_e}{m_\mu \sqrt{Z_L}}   \\
-\frac{y_{21}}{y_{11}}    \frac{m_e}{m_\mu \sqrt{Z_R}} & 1  \\
    \end{array}\).
\end{align}

$\Re{[\hat{C}_\g]}_{\mu\mu}$ is nothing but the coupling for the muon $g-2$ and~\cite{Crivellin:2013hpa}
\beq
\laq{g-2}
\frac{\Re[C_\g]_{\mu\mu}}{\L^2}=1.03\times 10^{-5}\TEV^{-2} \(\frac{\D a_{\mu}}{2.51\times 10^{-9}}\).
\eeq
On the other hand, the $\m \to e \g$ and electron EDM set bounds to the other components~\cite{Crivellin:2013hpa}
\beq
\frac{\sqrt{|(C_\g)_{e\m}|^2+|(C_\g)_{\m e}|^2}}{\L^2}\lesssim 2.1 \times 10^{-10}\TEV^{-2} \sqrt{\frac{{\rm Br}^{\rm bound}_{\m \to e \g}}{4.8\times 10^{-13}}}
\eeq
and 
\beq \frac{\Im{[C_\g]}_{ee}}{\L^2} \lesssim 5.3\times 10^{-12}\frac{|d_e|^{\rm bound}}{1.1\times 10^{-29} {\rm e cm}},\eeq respectively.
Here we have used the bounds obtained  by the MEG collaboration~\cite{TheMEG:2016wtm} and the ACME collaboration~\cite{Andreev:2018ayy} for the $\m \to e \g$ and electron EDM, respectively. 
We arrive at 
\beq
\laq{muegamma}
\sqrt{\frac{|y_{12}|^2}{y_{11}^2 Z_L}+\frac{|y_{21}|^2}{y_{11}^2 Z_R}} \frac{|C_\g|}{\Re{C_\g}} <0.00391363 \(\frac{2.7\times 10^{-9}}{\D a_{\mu}}\)\sqrt{\frac{\rm Br_{\mu \to e \g}^{\rm bound}}{4.8\times 10^{-13}}}
\eeq
and 
\beq
\laq{EDM}
\frac{\Im{[C_\g \frac{y_{12}y_{21}}{y^2_{11}} ]}}{\Re{[C_\g]}} \frac{1}{\sqrt{Z_L Z_R}} \lesssim 0.02 \(\frac{|d_e|^{\rm bound}}{1.1\times 10^{-29} {\rm e cm}}\) \(\frac{\D a_{\mu}}{2.51\times 10^{-9}}\).
\eeq 
To our model, $\m\to e \g$ constraint is most restrictive. A necessary condition to satisfy the constraint is
\beq\laq{neccon} \max{[\frac{|y_{12}|}{y_{11}\sqrt{Z_{L}}}, \frac{|y_{21}|}{y_{11}\sqrt{Z_{R}}}]} \lesssim 0.001-0.01.\eeq 
This requires a large $\sqrt{Z_R}, \sqrt{Z_L}$  or/and 
mild tunings on the off-diagonal components of $y_{ij}$ compared to the diagonal ones. This tuning may be related with UV physics. 
The suppression may also due to a large value of $\sqrt{Z_R }$ and $\sqrt{Z_L}$ if the BSM model is strongly coupled, e.g. in a conformal UV theory. 
We will discuss both cases based on the model in the next section.

In perturbation theory $\sqrt{Z_{L,R}}$ can be also large due to the RG running over a large hierarchy, $\L_{\rm cutoff} \gg \L. $ 
One obtain 
\beq
Z_{L,R}= \exp{[\int^{\L}_{\L_{\rm cutoff}}{ d\ln {\mu} \times \g_{L,R}}]}.
\eeq
where $\g_{L,R} $ are the anomalous dimension of the left handed and right-handed lepton of the  flavor ``2". 
Defining $\bar{\g}_{L,R}\equiv (\int^{\L}_{\L_{\rm cutoff}}{ d\ln {\mu} \times \g_{L,R}})/\log{(\L_{\rm cutoff}/\L)}$, this is 
\beq
\laq{form}
\sqrt{Z_{L,R}}\sim 400^{ \frac{\log{(\L_{\rm cutoff}/\L)}}{30}\cdot \frac{\bar{\g}_{L,R}}{0.4}}. 
\eeq
The anomalous dimension is a loop suppressed quantity, and $\bar{\g}_{L,R}= \O(0.1)$ 
requires a relatively strong coupling constant of $\O(1)$ in the UV theory. 
As we will see in the next section, a model can still satisfy the perturbative unitarity.

Since both the lepton flavor violation or CP-violation processes exist, 
this scenario may be probed via them even if the mass scales of the BSM particles are high.  
A model independent reach will be the regime with $\max{[\frac{|y_{12}|}{y_{11}\sqrt{Z_{L}}}, \frac{|y_{21}|}{y_{11}\sqrt{Z_{R}}}]}>10^{-4}-10^{-3} $ since the reach of the $\mu \to e \g$ is ${\rm Br}^{\rm bound}_{\m \to e \g}\sim 10^{-15}$ at MEG-II~\cite{Renga:2018fpd}. Given an $\O(1)$ CP phase of $C_\g$, the $\mu \to e \g$ rate and electron EDM have collation. The both detections will be a smoking gun signature of our scenario. 
Although it is absent for the model in the next section, our scenario allows the muon to have a large EDM which can be tested in the future by e.g. J-park~\cite{Gorringe:2015cma, Crivellin:2018qmi, Abe:2019thb}
Furthermore, all the $g-2$ scenarios can be fully tested in muon colliders~\cite{Capdevilla:2020qel, Buttazzo:2020eyl,Yin:2020afe, Capdevilla:2021rwo } even if the heavy particles are beyond the muon collider's reach~\cite{Yin:2020afe, Capdevilla:2021rwo}. 
In case the BSM particles are within the reach, a particular model may be confirmed by measuring the relevant cross-sections and decay channels of the BSM particles \cite{Yin:2020afe}.

\section{Lepton-slepton-bino-like system}

To be more concrete, we consider a simplified model similar to the lepton-slepton-bino system in the MSSM models (See Refs.~\cite{Endo:2013lva, Yamaguchi:2016oqz, Yin:2016shg, Yanagida:2018eho, Yanagida:2020jzy} for the muon $g-2$ explanations, and the both explanations of the muon and electron $g-2$~\cite{Endo:2019bcj} in this system). 
The purpose of this section is to study how large the radiatively induced lepton masses and $Z_L, Z_R$ could be in a perturbative model. 
We will also discuss extra-dimensional theory embedding the model with mild suppression of $y_{12,21}/y_{11}$.

The Lagrangian is given by 
\beq
\laq{Lag}
{\cal L_{\rm BSM}}= -g_L \bar{L}_2 \hat{P}_L \tl{L}_a \l^a-g_R \bar{e}_2 \hat{P}_R \tl{e}_a \l^a- A  H_0 \tl{L}_a^* \tl{e}^a-\frac{M_\l}{2} \bar{\l}_a \l^a+h.c.. 
\eeq
Here I define the Lagrangian with normalized kinetic terms at around the renormalization scale $\mu \sim \L$; $\l_a$ are $N$ copies of majorana fermion, ``bino"; $\tl{L}_a,\AND \tl{e}_a$ are $N$ copies of ``sleptons" of left-handed type and right-handed type, respectively; $g_{L,R}, A, M_\l$ are couplings and mass parameters. In addition, $m_{L,R}$ will be used for the mass of $\tl L$ and $\tl e$, respectively. 
The multiplicity of $N$ are introduced for later convenience. Each of $\tl{L}, \tl{e}, \l$  is a $N$ multiplet of an unbroken $Z_N$ symmetry or of $\SU(M)$ gauge symmetry
 with $M^2-1=N$, i.e. $N$ is the adjoint representation (which is the only choice for canceling the gauge anomaly).
Thus those particles can only couple to a single linear combination of the leptons.  
We perform a field redefinition to obtain positive $M_\l \AND A$. 
 In this basis, generically $g_L g_R^*$ has a non-vanishing CP phase.

By integrating out the BSM fields we obtain a radiatively induced Yukawa coupling at the one-loop level: 
\beq
\frac{\d Y}{\sqrt{Z_L Z_R}} = A N  \frac{g_L g_R^* M_\l }{16\pi^2} I(M_1^2, m_{L}^2, m_{R}^2). 
\eeq
Here the loop function is defined as
$
 I(x,y,z) = -\frac{xy\ln (x/y) + yz \ln (y/z) + zx \ln (z/x)}{(x-y)(y-z)(z-x)},
$
which satisfies $I(x,x,x)=1/2x$ and $I[x,x,0]\to 1/x$.

Assuming $M_\l\sim m_L\sim m_R \sim  A/\e \sim \L$ for simplicity, here and hereafter, we obtain
\beq
\laq{lep}
m_{\mu} \sim \frac{| \d Y|}{\sqrt{Z_L Z_R}} v \sim 0.1 \GEV \(\frac{\e N |g_L g_R^*| }{0.1}\) .
\eeq
We note that $\e<1$ is needed for the vacuum stability of the ``slepton" potential. 
In this case, it is easy to generate the lepton mass around $0.1-10\GEV.$
On the other hand, the higher dimensional photon coupling is estimated as~(c.f. \cite{Yin:2020afe})
\beq
\laq{Cg}
\frac{C_\g}{\sqrt{Z_L Z_R}\L^2} \sim
 \frac{1}{(300\TEV)^2}  \(\frac{ \e  N g_L g_R^* }{0.1}\) \(\frac{10\TEV}{ \L}\).
\eeq
Interestingly, in the effective theory, we can remove the CP phases of both $\d Y$ and $C_\g$ simultaneously by the field redefinition \cite{Borzumati:1999sp}. 
Thus in the following, we will take $g_L, g_R$ to be real positives to ease the analysis. 
However, we  still have the non-vanishing electron EDM from the phase of $y_{12} y_{21}.$
From \Eq{g-2}, surprisingly, the $g-2$ can be explained when the heavy physics mass scale to be even above $\O(10)\TEV$. 
This is the consequence of the absence of the chirality suppression, i.e. \eq{Cg} has no suppression of small $y_{ij}$. 
Also, the radiative correction to the Yukawa coupling is  not suppressed by the small coupling and indeed a Yukawa coupling much larger than that for the electron is radiatively induced by the same loop.

\Eqs{muegamma}  and \eq{EDM} can be satisfied if \eq{neccon} is satisfied. 
To check this, let us estimate $Z_{L,R}$. 
In the renormalization scale $\mu \gtrsim \L$, the anomalous dimension for $L_2 (e_2)$ is
\beq
\g_{L(R)} \sim -\frac{3N g_{L(R)}^2}{16\pi^2}
\eeq
by neglecting other couplings than $g_L, g_R$. 
Also the beta function of $g_{L,R}$ can be obtained as 
\beq
 \frac{d g_{L}}{d \log{\mu}} \sim\frac{g_L }{16\pi^2} (\frac{3N -1}{2}g_L^2  +3g_R^2), ~~~~ \frac{d g_{R}}{d \log{\mu}} \sim\frac{g_R }{16\pi^2} g_R \frac{3g_L^2+(3N+2)g_R^2}{2}
\eeq 
by  again neglecting other couplings. (We will take account of them later.) 
One can solve the RGE to calculate $g_L$ and $g_R$ and estimate $Z_L$ and $Z_R$.  

To solve the RGE, I take  $\L_{\rm cutoff}$ to be the scale at which $g_L= g_R=\sqrt{4\pi}.$\footnote{$g_L=g_R$ is a (pseudo) IR fixed point, representing an approximate $\SO(3)$ symmetry.  }
This should give the largest $Z_{L,R}$ for a given $\L_{\rm cutoff}/\L.$ 
 Fig.\ref{fig:res} displays $\sqrt{Z_{L}}(=\sqrt{Z_R})$ with $\L_{\rm cutoff}/\L\approx 10^{13},\AND 2.2\times 10^4$ for red solid and dashed lines, respectively.  We also show the corresponding  $N g_L g_R$ at the renormalization scale $\mu =\L$. This quantity is important for both the induced lepton mass \eq{lep} and the photon coupling \eq{Cg}.
One can see that the lepton mass can be as large as a few $\GEV$ with $\L_{\rm cutoff}$ in an intermediate scale or/and $N$ is large. 
The $\sqrt{Z_{L,R}}$ suppression can be as large as $\O(10-100)$ for $N=10-10^3.$ Such a large multiplicity may result from an $\SU(M)$ gauge theory. For instance, the adjoint representation of $\SU(5)$ has $N=24.$ However, $N\gg 100$ may not be natural, and the 1loop analysis may not be valid.
Thus, to evade the $\mu\to e \g$ bound \eq{muegamma}, we need a mild-tuning of $\max{[|y_{12}|/y_{11},|y_{21}|/y_{11}]}< \O(10)\%$ unless we take account of other coupling effects (see the following).
The $\O(1-10)\%$ tuning region can be tested in the MEG-II experiment.

$y_{12}/y_{11}$ can also be mildly suppressed by UV dynamics. 
For instance, we can consider that a set of lepton $L_{\rm bulk}, e_{\rm bulk}$, $H_0$ live in a bulk in a 5D model of orbifold $S_1/Z_2$ at high energy. 
Then from gauge invariance, the SM gauge bosons (of  $\SU(2)_L\times \U(1)_Y$) also live in the bulk. We assume that there is also Yukawa interaction among those particles in the bulk.
The other leptons and $\l, \tl{e},\tl{L}$ are supposed to live on a brane at the orbifold singularity. 
This means the slepton-lepton-bino interaction is localized on the brane. 
Then one finds at the compactification scale, $1/R$, of the 4D effective theory that  the 
 bulk fields acquire wave functions of $\sim R \times \L_5$, with $\L_5$ being the 5D cutoff scale, which we take to be universal for all bulk fields, $R$ the volume of the extra dimension, which is related to the 4D cutoff scale, $R\sim 1/\L_{\rm cutoff}$. 
Performing a compactification at $\mu \sim 1/R$ and normalizing the kinetic term, 
we find that 
the brane leptons are $\sim L_2,e_2$ and the bulk leptons are $L_{\rm bulk}\sim L_1, e_{\rm bulk}\sim e_1$ since the bino-bulk [brane] lepton-bulk slepton interaction is suppressed by $(R\L_5)^{-1/2} [(R\L_5)^{0}].$
One also gets $y_{12}, y_{21}\propto (R \L_5)^{-1/2}, y_{11},y_{22}\propto (R \L_5)^{0}.$
This means $y_{12}/y_{11},y_{21}/y_{11}$ is naturally suppressed by $(R\L_5)^{-1/2}$. 
For the 5D gauge coupling to be weakly coupled at $\m \sim \L_5$, we need $\L_5 R \lesssim \frac{24\pi^2}{g_2^2(1/R)}\sim 10^3.$\footnote{Due to this constraint, one may not explain the $m_e/m_\mu$ hierarchy purely with the volume suppression. If a certain hierarchy of the Yukawa couplings is set at $\L_5$ to explain the SM lepton mass and suppose the BSM fields to explain the $g-2$ without inducing a radiative lepton mass, 
there are tunings for the alignment between the Yukawa interaction basis and the brane field. }
Thus the $|y_{12,21}/y_{11}| \gtrsim 10^{-2}\text{-}10^{-1}$ can be realized in the perturbative regime naturally. This UV model can  successfully explain the muon mass as well as the $g-2$ with $N= \O(1)$. 
\\

In the case of the introduction of the $\SU(M)$ gauge group or when the SM gauge coupling become strong we would have other contribution to modify the RG running.\footnote{In particular, we need to  consider the effect when $\SU(M)$ is the SM gauge group, such as the color group $\SU(3)_c.$ 
In this case $\l_a$ could be identified as the PQ fermion whose mass is from the PQ symmetry breaking. As a result, the strong CP problem can be solved by the QCD axion. The axion may be even heavier than the conventional one if $N$ is large and sleptons are lighter than the binos (see the model in the appendix  of \cite{Kitano:2021fdl}).
}
At the one-loop level, on the other hand, the form of $\g_{L,R}$ does not change. Since the gauge contributions slow down the runnings of $g_L, g_R$, 
we can estimate the maximized $\sqrt{Z_L}$ by taking  $g_{L,R}= \sqrt{4\pi}$ at any scale. 
Then we obtain 
\beq
\bar{\g}^{\rm max}_{L,R}= 0.24 N
\eeq
in perturbation theory. In this case from \eq{form} we can have $\sqrt{Z_{L,R}}\gg 10-100.$


\begin{figure}[t!]
\begin{center}  
\includegraphics[width=105mm]{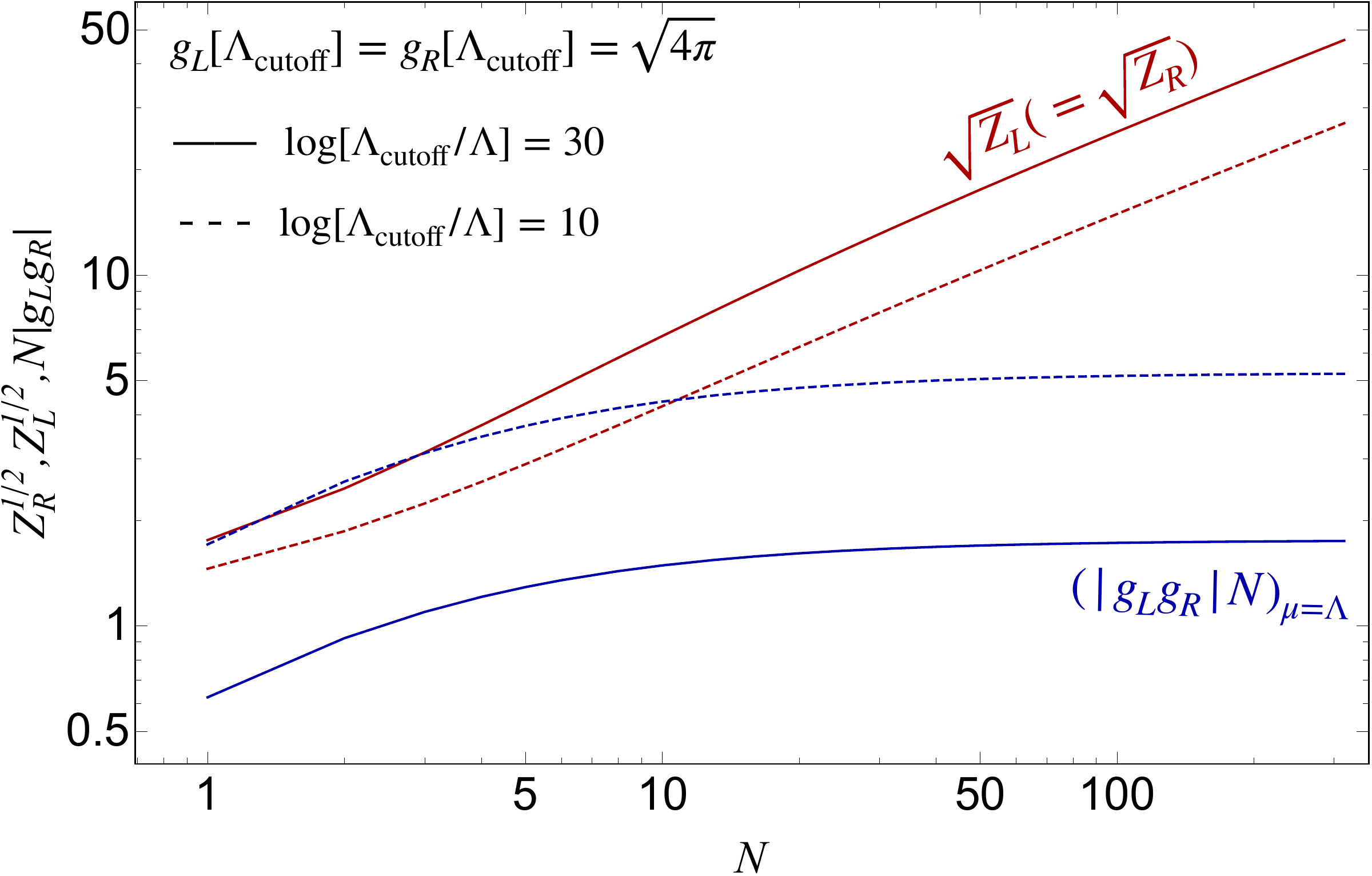}
\end{center}
\caption{The $\sqrt{Z_L}(=\sqrt{ Z_R})$ (red lines) and $|g_L g_R| N$ at $\mu=\Lambda$ (blue lines) by varying the multiplicity $N$. The solid (dashed) lines are for the hierarchy $\log{\L_{\rm cutoff}/\L}=30 \AND 10$ respectively. 
}
\label{fig:res}
\end{figure}

Before ending this section, let me mention that the symmetry, for the multiplicity $N$, stabilizes the lightest BSM particles. In particular $\l$ should be a good candidate for dark matter. If $\L$ is not very large, the dark matter can annihilate into the heavy leptons  in the early and present universe. 
This is because the lepton coupled to the dark matter automatically becomes the heavy lepton due to the dark matter loop. 
Thus, we may have the WIMP miracle, and the signal of heavy lepton pair in the indirect detection experiment may be a good probe of this scenario.

\section{Discussion and conclusions }

\paragraph{Tau lepton mass and phenomena}

So far we have focused on the simplified two flavor model. It is straightforward to extend to the three flavor cases. We may introduce another ``family" of the BSM particles, e.g.  $\tl{e}_3\AND \tl{L}_3 $ in the context of the concrete mode, to generate the tau lepton mass. It can be seen to be viable in Fig.\ref{fig:res} to generate an $\O(1)\GEV$ fermion mass, especially when the typical mass scale, $\L'$, of the sleptons are closer to 
$\L_{\rm cutoff}.$ 
To check the flavor and CP violations we can first take the Yukawa $y_{ij}=\O(m_\mu/v)$ for the $i, j=2\OR 3$ otherwise $m_e/v$ (i.e. we have integrated out the family of sleptons, $\tl{L},\tl{e}$, for muon.) 
Here $L_3 e_3$ are the leptons coupled to the BSM particles for tau mass. 
 By noting that the photon coupling scales with $\L'$,  but the induced Yukawa coupling does not, one can easily evade the bound for $\tau$ lepton with large $\L'.$ 
On the other hand, if $\L'\sim \L$, we get the $g-2$ of the tau lepton, $|\D a_\tau| \sim (m_\t/m_\mu)^2\times \D a_{\mu}\sim 10^{-7}$, 
but, perhaps, the tau $g-2$ is difficult to be tested in the near future e.g. \cite{Beresford:2019gww}. 
The rate of $\tau\to \mu \g$  is around the current bound and this may be tested.\footnote{As we can expect, the electron mass can be also induced radiatively by further introducing a weakly coupled family of the BSM particles without spoiling the suppression mechanism. In this radiative generation of the Yukawa couplings, the hierarchy of $(g_{L} \cdot g^*_{R})_\a$ becomes the hierarchy of the SM Yukawa couplings. In terms of $g_L \sim g_R$ the hierarchy becomes milder $ \sqrt{m_e}: \sqrt{m_\mu}: \sqrt{m_\tau}.$ }

\paragraph{Suppression of $y_{ij}$}
So far we have assumed the Yukawa coupling in the BSM model satisfying  $y\sim m_e/v \ll 1$ which is small. 
However,  the small electron mass may be chosen by an anthropic selection~\cite{Agrawal:1997gf}. 
Therefore given that $y_{ij}\sim y$,  $y_{ij}$ may need to be all small due to an anthropic reason. 

Alternatively, volume suppression of more complicated manifolds than $S_1/Z_2$ in extra-dimension may also explain the small $y_{11,22}$ and even smaller $y_{12,21}$.

\paragraph{Application to B-physics anomaly}
One can also apply the mechanism to scenarios explaining other anomalies, like the recently reported B-physics anomaly suggesting a lepton non-universality~\cite{Aaij:2021vac}. 
Again the BSM generate such non-universality should not induce too large $\mu \to e \g$ rate and the electron EDM. 
Thus the result also implies the flavor/CP structure of the BSM. 

For instance, the lepton non-universality can be explained by a leptoquark~e.g. \cite{Bauer:2015knc,ColuccioLeskow:2016dox} with Lagrangian 
\beq
{\cal L}= (\bar{Q^c} \l_L  i \tau_2 L  +\bar{u}^c \l_R e) \F
\eeq
where $\l_L, \AND \l_R$ are Yukawa matrices, and $\F$ is the leptoquark with charge $\bar{3}, 1, \AND 1/3$ under the representation of $\SU(3)_c, \SU(2)_L, \AND \U(1)_Y$. 
This is nothing but the Lagrangian \eq{Lag} with the replacement $\tl{L}, \tl{e},\l \to \bar{Q}^c, \bar{u}^c, \F$. 
Therefore with a general flavor structure, the model should be constrained as well by the electron EDM  and $\mu \to e \gamma$ induced by quark-leptoquark loop. 
Here we propose that if the heavy lepton masses are generated through the quark-leptoquark loop proportional to quark Yukawa couplings, the 
higher dimensional terms are aligned to the heavy lepton masses. As a consequence, the lepton flavor and CP violations relevant to the electron are suppressed.  
On the other hand, the quark masses are also induced with lepton loops, but it is suppressed by $y_{22}\sim m_e/v$ and is negligible. 

\paragraph{Conclusions}

The large muon $g-2$  but suppressed $\m \to e \g$ and electron EDM is a hint of the flavor and CP structure of the BSM. 
Here I studied that the $g-2$ explanation associated with the generation of  masses of the heavy SM leptons. I showed that in this case, the $\m \to e \g$ rate and electron EDM are naturally suppressed since the basis of the higher dimensional operators and masses are automatically aligned. However, they are non-vanishing and can give experimental signals. 
  The further measurement of the $g-2$ together with flavor and CP violation may reveal the origin of the heavy lepton masses. 
  \\

{\it Note added:} While completing the first version of this paper, I found Ref.~\cite{Baker:2021yli}, in which the authors classify the minimal models for both generating a one-loop radiative mass of the muon and the $g-2$. 
On the contrary, we show that the lepton flavor violation and electron EDM are generically suppressed when the $g-2$ operator and muon mass are radiatively induced at the same time.

\section*{Acknowledgements} 
The author thanks Tomohiro Abe and Fuminobu Takahashi for useful discussions. 
The author is also grateful to the University of Tokyo High Energy Physics Theory Group for many bits of help during the COVID epoch when he was a member of the group. 
This work was supported by JSPS KAKENHI Grant-in-Aid for Scientific
Research 19H05810 and 20H05851.

\providecommand{\href}[2]{#2}\begingroup\raggedright\endgroup
\end{document}